\pdfoutput=1
\documentclass[a4paper,11pt]{article}

\usepackage[english]{babel}
\usepackage[utf8]{inputenc}
\usepackage[T1]{fontenc}
\usepackage[normalem]{ulem}

\newif\ifanonymous

\usepackage[a4paper,top=2cm,bottom=2cm,left=2cm,right=2cm,marginparwidth=1.75cm]{geometry}

\usepackage{amsmath, amssymb, amsthm}

\usepackage{thmtools}
\usepackage{mathtools}
\usepackage[boxed]{algorithm2e}
\usepackage{todonotes}
\usepackage{placeins}
\usepackage{subcaption}
\usepackage{booktabs}
\usepackage{thm-restate}
\usepackage{graphicx}
\definecolor{VUBlue}{rgb}{0,0.466,0.702}
\usepackage{cite}
\usepackage[colorlinks=true, allcolors=VUBlue]{hyperref}
\usepackage{doi}
\usepackage[capitalize]{cleveref}
\usepackage{xspace}
\usepackage{newtxtext}
\usepackage{newtxmath}
\usepackage{tikzsymbols}
\usepackage{dirtytalk}
\usepackage{microtype}

\usepackage{tikz}
\usetikzlibrary{arrows.meta}

\newcommand{\algstep}[1]{%
  \vspace{2pt}%
  \textcolor{VUBlue}{\bfseries $\blacktriangleright$\ #1}\;
  \vspace{1pt}}

\definecolor{CWIBlue}   {rgb}{0.38, 0.478, 0.56}
\definecolor{CWIDGreen} {rgb}{0.372,0.494, 0.419}
\definecolor{CWILGreen} {rgb}{0.603,0.717, 0.486}
\definecolor{CWIRed}    {rgb}{0.686,0.188, 0.262}
\definecolor{VUBlue}    {rgb}{0,    0.466, 0.702}
\definecolor{VUBlue2}   {rgb}{0.364,0.678, 0.815}
\definecolor{VUGreen}   {rgb}{0,    0.502, 0.325}
\definecolor{VUGreen2}  {rgb}{0.309,0.686, 0.282}
\definecolor{VUOrange}  {rgb}{0.8,  0.254, 0}
\definecolor{VUOrange2} {rgb}{0.909,0.411, 0.176}
\definecolor{VUPurple}  {rgb}{0.231,0.129, 0.443}
\definecolor{VUPurple2} {rgb}{0.556,0.302, 0.643}

\newcommand{\LBIF}{\textsf{LBIF}\xspace}

\newcommand{\ie}[1]{(i.e.,~#1)}
\newcommand{\eg}[1]{(e.g.,~#1)}
\newcommand{\largerm}{\mathrm{large}}
\newcommand{\smallrm}{\mathrm{small}}
\newcommand{\meanLoad}{\ell}
\newcommand{\capa}[1]{c_{#1}}
\newcommand{\capas}{\mathcal{C}}
\newcommand{\ites}{\mathcal{S}}
\newcommand{\ite}[1]{s_{#1}}
\newcommand{\iteSmall}{\ite{\smallrm}}
\newcommand{\iteLarge}{\ite{\largerm}}
\newcommand{\itesSmall}{\ites_{\smallrm}}
\newcommand{\itesLarge}{\ites_{\largerm}}
\newcommand{\totSize}{W}
\newcommand{\numBins}{B}
\newcommand{\targetLoad}{target load\xspace}
\newcommand{\targetLoads}{target loads\xspace}

\newif\ifhidetodos

\usepackage{marginnote}

\newcommand{\cbox}[2][yellow]{%
  \fcolorbox{#1}{white}{\parbox{\dimexpr\linewidth-2\fboxsep}{\strut #2\strut}}%
}

\newcommand{\customtodo}[3]{\textcolor{#2}{\(\bla\meanLoadktriangledown\)}\marginnote{\raggedright \textcolor{#2}{\textbf{#1:} #3}}}
\newcommand{\customtododisplay}[3]{\noindent\cbox[#2]{\textcolor{#2}{\textbf{#1:} #3}}}
\newcommand{\customtodoinline}[3]{\textcolor{#2}{\textcolor{#2}{\textbf{#1:} #3}}}

\ifhidetodos
    \renewcommand{\customtodo}[3]{}
    \renewcommand{\customtododisplay}[3]{}
    \renewcommand{\customtodoinline}[3]{}
\fi

\definecolor{TFuglypink}{rgb}{1,0.41,0.71}

\newcommand{\cO}{\mathcal{O}}
\newcommand{\cOPT}{\mathrm{OPT}}
 \newcommand{\defproblem}[3]{
  \vspace{2mm}
\noindent\fbox{
  \begin{minipage}{0.96\columnwidth}
  \textsc{#1}\\
  {\bf{Input:}}\ #2  \\
  {\bf{Output:}} #3
  \end{minipage}
  }
  \vspace{2mm}
}

\title{Load Balancing with Item Fragmentation and \\ More Bins than Items: Linear-Time and Exact}

\ifanonymous
    \author{Anonymous Author(s)}
\else
    \usepackage{authblk}
    \usepackage{orcidlink}

\author[1,2]{Ben Bals \orcidlink{0009-0009-1054-8444}}
\author[3]{Tom\'{a}\v{s} Flouri \orcidlink{0000-0002-8474-9507}}
\author[4]{Kassian Kobert \orcidlink{0009-0001-9930-8739}}
\author[5]{Solon P. Pissis \orcidlink{0000-0002-1445-1932}}
\author[1,2]{Leen Stougie \orcidlink{0000-0001-6938-8902}}
\affil[1]{CWI, Amsterdam, The Netherlands}
\affil[2]{Vrije Universiteit, Amsterdam, The Netherlands}
\affil[3]{University College London, UK}
\affil[4]{Universität Bielefeld, Germany}
\affil[5]{The Cyprus Institute, Nicosia, Cyprus}
\fi

\declaretheorem[numberwithin=section]{theorem}

\declaretheorem[sibling=theorem]{corollary}
\declaretheorem[sibling=theorem]{lemma}

\declaretheorem[sibling=theorem]{observation}

\declaretheorem[name=Invariant]{invariant}
\Crefname{invariant}{Invariant}{Invariants}
\Crefname{observation}{Observation}{Observations}

\begin{document}
\maketitle

\begin{abstract}
We study a fundamental load-balancing problem in which $N$ items of strictly positive integer sizes with total size $\totSize$ must be assigned to $\numBins$ bins, with items allowed to be divided into integer-size \emph{fragments} across bins. Each bin must receive a load of either $\lfloor \totSize/\numBins \rfloor$ or $\lceil \totSize/\numBins \rceil$, ensuring that the loads of any two bins differ by at most one unit. Under this  constraint, the objective is to minimize the maximum number of distinct item fragments placed into any single bin. This problem is NP-hard and there exists an $(\mathrm{OPT}+1)$-approximation algorithm that runs in $\cO(N \log N + \numBins)$ time [Kobert et al., 2014]. The existing algorithm is employed by several widely used production codes for phylogenetic reconstruction.

Here, we focus on the case $\numBins \ge N$, that is, bins are at least as numerous as items. 
This setting arises pervasively in modern parallel and distributed systems, as well as in iterative algorithms that rebalance their workload at every step. In this context, bins are machines or machine cores, and items represent computational tasks (jobs).
Our central contribution is a remarkably simple $\cO(N)$-time algorithm that computes an optimal assignment in this setting; this assignment is encoded using $\cO(N)$ space. In particular, we prove that an optimum always exists in which every bin receives \emph{at most two fragments}
and provide a characterization of when one fragment per bin suffices. 
Our result strictly improves the $(\mathrm{OPT}+1)$-approximation of Kobert et al.~in the many-bins setting \emph{and}, together with it, yields a \emph{strongly} polynomial-time $(\mathrm{OPT}+1)$-approximation algorithm for the general setting, as it removes the dependency on the number $\numBins$ of bins.
\end{abstract}

\newpage

\section{Introduction}

Balancing work across processing units is a fundamental problem in parallel and distributed computing,
from early multiprocessors~\cite{graham1969bounds,cybenko1989loadbalancing} to modern many-core systems~\cite{shiina2019workstealing,los2022balanced}.
As processor core counts grow, efficient utilization becomes increasingly important:
a system draws substantial power whether its cores are busy or idle~\cite{barroso2007energy}, so unused cores
waste capacity, whereas shorter runtimes reduce total
energy use~\cite{woo2008extending}. The problem is becoming more acute
as datasets grow across disciplines, increasing the work to be
balanced: from large-scale simulations in high-performance and scientific
computing~\cite{karypis1998metis,blazewicz1997startup,veeravalli2003dlt}, through
web-scale queries and MapReduce jobs in data
systems~\cite{dean2008mapreduce,gufler2011skew,kolb2012loadbalancing} and rising 
traffic in communication networks~\cite{krysta2003splittable,correa2015splittable},
to billion-parameter models shared across accelerators in machine
learning~\cite{li2014paramserver,mudigere2022dlrm}. In such settings, the
workload is often divisible and should be distributed evenly, while each processing unit may incur an overhead for
every distinct (potentially fragmented) work unit it handles, motivating us to minimize the number of distinct units.

This led us to study the following problem, which has been introduced and studied in~\cite{DBLP:conf/wabi/KobertFAS14}.
Given $N$ items (jobs) of strictly positive integer sizes and total size $\totSize$, assign them to $\numBins$ bins (processors), with items allowed to be divided into integer-size \emph{fragments} across bins.
The load between bins must be as balanced as possible, that is, a load of either $\lfloor \totSize/\numBins \rfloor$ or $\lceil \totSize/\numBins \rceil$.
Under this  constraint, the objective is to minimize the maximum number of distinct item fragments placed into any single bin.

A concrete  application of balancing divisible load while limiting per-processor
fragmentation arises in statistical phylogenomics~\cite{kapli2020review}.
The input is a multiple sequence alignment, which can be viewed as a matrix:
its rows correspond to taxa and its columns to homologous positions in their
molecular sequences. Each column is called an \emph{alignment site}.
The alignment sites are typically grouped into \emph{partitions}, with the
sites in each partition analyzed under the same evolutionary model.
For a fixed candidate phylogenetic tree, likelihood calculations at different sites are
mutually independent, and their repeated evaluation dominates the cost of
maximum-likelihood and Bayesian inference~\cite{felsenstein1981}. Alignment sites can
therefore be distributed across processors. Under the standard uniform-site
cost model, assigning approximately the same number of sites to every processor
balances the arithmetic workload. However, a processor also incurs
partition-specific overhead for every partition from which it receives at least
one site. The goal is thus to balance the number of sites while
minimizing the maximum number of distinct partitions handled by any processor.
Kobert et al.~\cite{DBLP:conf/wabi/KobertFAS14} formalized this task as the
\emph{divisible load balance problem}, which we will call here Load Balancing with Item
Fragmentation (\LBIF). In this formulation, the items are the partitions, the
size of an item is its number of sites, and the bins are the processors. A
partition may be divided among several processors, and each processor receives
either $\lfloor W/B\rfloor$ or $\lceil W/B\rceil$ sites, where $W$ is the total
number of sites and $B$ is the number of processors. The objective is to
minimize the maximum number of partitions assigned to any one processor.
Kobert et al.~proved the problem to be NP-hard and gave an
$(\mathrm{OPT}+1)$-approximation running in $\cO(N\log N+B)$ time for $N$
partitions (items). Their algorithm is implemented in \emph{widely used} phylogenetic software,
including RAxML-NG~\cite{kozlov2019raxmlng} and ExaBayes~\cite{aberer2014exabayes}.
A related, data-dependent variant arises from the site-repeats
optimization~\cite{kobert2017siterepeats}. For a node of the
phylogenetic tree, a site's \emph{subcolumn} is its alignment column restricted to the
taxa in the subtree rooted at that node. 
If two sites in the same partition have identical subcolumns, the intermediate
likelihood at the root of that subtree is the same for both and needs to be
computed only once. Consequently, workload is not determined by site count alone,
and gives another variation of the load-balancing problem~\cite{scholl2016,morel2017siterepeats}.

In this paper, we study the case in which there are at least as many bins as
items. In practice, bins may far outnumber items: a modern server may provide
hundreds of cores, and a cluster many more, while potentially only a handful of large,
divisible jobs are active at a time. The algorithm of Kobert et al.~applies in
this regime, but remains an approximation: it returns an assignment of value at most
$\mathrm{OPT}+1$ and does not necessarily attain $\mathrm{OPT}$. Moreover, its
$\cO(N\log N+B)$ running time depends on the number $B$ of bins. Since $B$ may be
arbitrarily large relative to $N$, this bound may not be polynomial in $N$.
In response, we show that, for every $B\geq N$, an \emph{optimal} assignment in which
every bin receives at most two fragments always exists. Furthermore, we construct such an
assignment in $\cO(N)$ time using $\cO(N)$ space, and give an $\cO(N)$-time
arithmetic test that determines whether the optimum is one or two. Combined
with the algorithm of Kobert et al.~for $B<N$, our result yields an
$(\mathrm{OPT}+1)$-approximation for every $B$ that runs in
$\cO(N\log N)$ time and is therefore strongly polynomial.
Additionally, our algorithm is so simple that it is easy to implement, making it amenable to practical adoption \eg{in the existing software~\cite{aberer2014exabayes,kozlov2019raxmlng}}.
The $\cO(N)$ running time is also beneficial when assignments must be recomputed
frequently. This occurs, for example, in fault-tolerant likelihood evaluation,
where data are redistributed whenever processors fail or become
available~\cite{huebner2021faulttolerance}.
Another application arises in Metropolis-coupled MCMC for Bayesian
phylogenetic inference, which runs several interacting chains at different
temperatures in parallel~\cite{altekar2004}. When the available cores outnumber
the chains, the likelihood computation within each chain can be divided across
independent alignment sites. Treating chains as divisible items
and cores as bins yields a many-bins instance in which per-step work is balanced
while the maximum number of chains handled by any core is minimized.

\paragraph{Related work.}
Apart from the above-mentioned work motivating our study, we mention here literature on multiprocessor scheduling and bin packing. Load balancing is the most common target in classical multiprocessor scheduling, often under the name of \emph{makespan minimization}. Indeed, the first theoretical results date back to the sixties~\cite{graham1966bounds,graham1969bounds}. The general problem on identical machines is NP-hard~\cite{karp}. However, preemptive jobs make the problem easy~\cite{MacNaugton59}, and splittable (divisible) jobs, dividing jobs across identical machines without further constraints makes the problem trivial. Our problem is complicated by the objective of minimizing the maximum number of  
job fragments. 

Other works on \emph{splittable (divisible) scheduling}~\cite{krysta2003splittable,epstein2006onlinesplittable,correa2015splittable,jansen2018batch}
divide jobs into fragments across identical or unrelated machines to minimize the makespan, but with
the number of fragments bounded~\cite{krysta2003splittable,epstein2006onlinesplittable} or penalized, for instance, through (batch) setup
times~\cite{correa2015splittable,jansen2018batch}. 
As its name suggests, there is also a relation to \emph{bin packing with item
fragmentation}. In early work on this problem~\cite{menakerman2001bpif,shachnai2008aptas,shachnai2007fptas}, as well as in later publications~\cite{bertazzi2019bpif, casazza2014bpif,
ekici2021bpifconflicts}, fragmentation comes at a price: increasing cost or size. 

None of these related results are applicable in our setting, which fixes an exact load for every bin and minimizes the maximum number of fragments per bin, rather than the number of
bins or the makespan.

\subsection{Formal Problem Definition and Our Contribution}

We study the following fundamental load-balancing problem introduced in~\cite{DBLP:conf/wabi/KobertFAS14} by Kobert et al.

\defproblem{Load Balancing with Item Fragmentation (\LBIF)}
{A multiset $\ites = \{\ite{1}, \ite{2}, \dots, \ite{N}\}$, with $\ite{i}\in \mathbb{N}$, for all $1\leq i \leq N$, and an integer $\numBins>0$.}
{The minimum value of $\max_{1\leq j \leq \numBins} |\{i : x_{i,j} > 0\}|$ over all matrices $X = (x_{i,j}) \in \mathbb{N}_{\ge 0}^{N \times \numBins}$ subject to:
\begin{enumerate}
    \item $\sum_{j=1}^\numBins x_{i,j} = \ite{i}$, for all $1 \le i \le N$.
  \item $\sum_{i=1}^N x_{i,j} \in \{\lfloor \totSize/\numBins \rfloor, \lceil \totSize/\numBins \rceil\}$, for all $1 \le j \le \numBins$, where $\totSize = \sum_{i=1}^N \ite{i}$.
\end{enumerate}}

We refer to the integers $\sum_{i=1}^N x_{i,j}$ and $\max_{1
\leq j \leq \numBins} 
 |\{i : x_{i,j} > 0\}|$ as the \emph{load} and the \emph{combinatorial load}, respectively. 
 
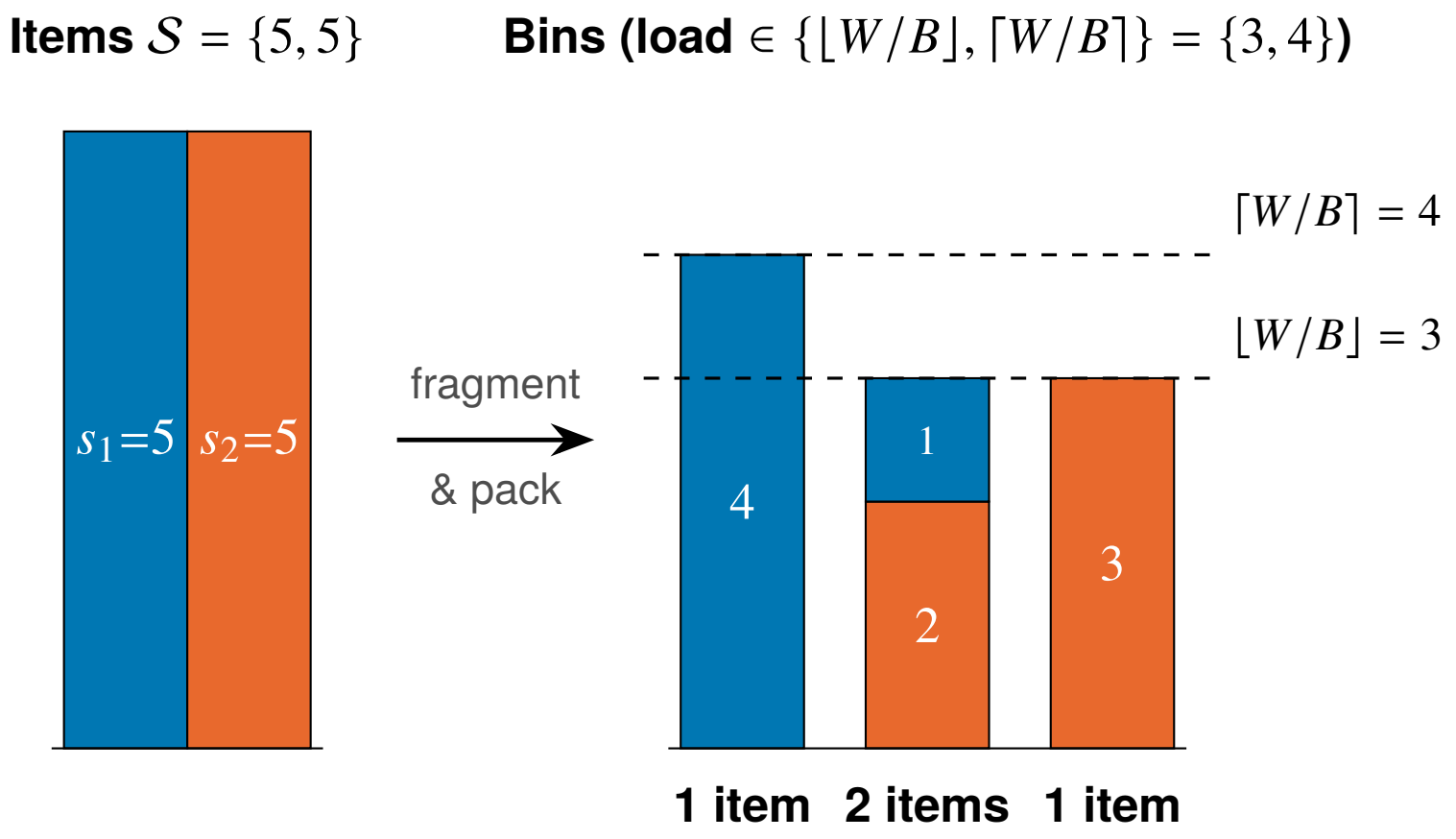
\begin{figure}[h]
\centering
\begin{tikzpicture}[
bar/.style ={draw=black, line width=0.4pt},
lbl/.style ={font=\small\sffamily, text=white},
hdr/.style ={font=\small\sffamily\bfseries},
captionStyle/.style ={font=\footnotesize\sffamily, text=black!70},
x=0.85cm, y=0.85cm
]

\node[hdr, anchor=south] at (1, 5.4) {Items $\mathcal{S}=\{5,5\}$};

\filldraw[bar, fill=VUBlue]    (0,0) rectangle (1,5);
\node[lbl] at (0.5,2.5) {$s_1{=}5$};
\filldraw[bar, fill=VUOrange2] (1,0) rectangle (2,5);
\node[lbl] at (1.5,2.5) {$s_2{=}5$};

\draw[black] (-0.1,0) -- (2.1,0);

\draw[-{Stealth[length=3mm]}, line width=0.9pt, black]
(2.7,2.5) -- (4.3,2.5);
\node[captionStyle, anchor=south] at (3.5,2.6) {fragment};
\node[captionStyle, anchor=north] at (3.5,2.4) {\& pack};

\node[hdr, anchor=south] at (7, 5.4)
{Bins  (load $\in\{\lfloor W/B\rfloor,\lceil W/B\rceil\}=\{3,4\}$)};

\filldraw[bar, fill=VUBlue]   (5,0) rectangle (6,4);
\node[lbl] at (5.5,2) {$4$};
\node[hdr, anchor=north] at (5.5,-0.15) {1 item};

\filldraw[bar, fill=VUOrange2] (6.5,0) rectangle (7.5,2);
\node[lbl] at (7,1) {$2$};
\filldraw[bar, fill=VUBlue]    (6.5,2) rectangle (7.5,3);
\node[lbl] at (7,2.5) {\scriptsize $1$};
\node[hdr, anchor=north] at (7,-0.15) {2 items};

\filldraw[bar, fill=VUOrange2] (8,0) rectangle (9,3);
\node[lbl] at (8.5,1.5) {$3$};
\node[hdr, anchor=north] at (8.5,-0.15) {1 item};

\draw[dashed, black, line width=0.6pt]
(4.7,3) -- (9.3,3)
node[anchor=south west, font=\footnotesize\sffamily, text=black]
{$\lfloor W/B\rfloor=3$};
\draw[dashed, black, line width=0.6pt]
(4.7,4) -- (9.3,4)
node[anchor=south west, font=\footnotesize\sffamily, text=black]
{$\lceil W/B\rceil=4$};

\draw[black] (4.9,0) -- (9.1,0);

\end{tikzpicture}
\caption{A simple example of an \LBIF instance with $N=2$ items and $B=3$ bins and a solution with optimal combinatorial load $2$: bin~$2$ contains fragments of both items.} \label{fig:example-load2}
\end{figure}

To explain the \LBIF problem informally and to streamline the presentation, we associate each element of the multiset $\ites$ with an \emph{item} and the integer $\numBins$ with the number of available \emph{bins}. We are given $N$ items of strictly positive integer size with total size $\totSize$. We must assign the items to $\numBins$ bins, where items may be fragmented across bins. Each bin must receive a load of either $\lfloor \totSize/\numBins \rfloor$ or $\lceil \totSize/\numBins \rceil$, ensuring that the loads of any two bins differ by at most one unit. Under this balancing constraint, the objective is to minimize the maximum number of item fragments placed into any single bin. 
See \Cref{fig:example-load2} for an example.
For simplicity of presentation, we use $\ite{i}$ both for the item index $i$ and for its size if the meaning is clear from context. Otherwise, we use the index $i$ explicitly to avoid confusion.

The following results are known for \LBIF.

\begin{theorem}[Kobert et al.~\cite{DBLP:conf/wabi/KobertFAS14}]
    \LBIF is NP-hard.
\end{theorem}

\begin{theorem}[Kobert et al.~\cite{DBLP:conf/wabi/KobertFAS14}]\label{the:known-UB}
There exists an $(\mathrm{OPT}+1)$-approximation algorithm for \LBIF that runs in $\cO(N \log N + \numBins)$ time.
\end{theorem}

    From a theory standpoint, one would like a \emph{strongly} polynomial-time approximation algorithm for \LBIF. When $\numBins=N^{\cO(1)}$, the $\cO(N \log N + \numBins)$-time $(\mathrm{OPT}+1)$-approximation algorithm of Kobert et al.~\cite{DBLP:conf/wabi/KobertFAS14} runs in polynomial $N^{\cO(1)}$ time. In particular, when $\numBins\leq N \log N$, their algorithm runs in $\cO(N \log N)$ time. We are thus motivated to study the regime $\numBins\geq N$. 
    We denote the set of \LBIF instances with $\numBins\geq N$ by $\LBIF_{N\le \numBins}$.
    
    Our central contribution is an \emph{exact} algorithm for $\LBIF_{N\le \numBins}$ that runs in the optimal $\cO(N)$ time. 
    Such a solution always has a combinatorial load of at most $2$.

\begin{theorem}\label{the:main}
There exists an algorithm for $\LBIF_{N\le \numBins}$ that runs in $\cO(N)$ time and always returns an exact solution with combinatorial load at most 2. In particular, the assignment is output in a representation using $\cO(N)$ space.  
\end{theorem}

Proving \Cref{the:main} will be the focus of this paper. Combined with \Cref{the:known-UB}, \Cref{the:main} yields a strongly polynomial-time approximation algorithm for \LBIF.

\begin{corollary}\label{cor:strong-pol}
There exists an $(\mathrm{OPT}+1)$-approximation algorithm for \LBIF that runs in $\cO(N \log N)$ time.
\end{corollary}
\begin{proof}
    After reading the $N$ item sizes, we compare $N$ with $\numBins$ in $\cO(1)$ time and run exactly one of the following two algorithms on the unchanged input.
    When $\numBins<N$, apply \Cref{the:known-UB}; its running time $\cO(N\log N+\numBins)$ is $\cO(N\log N)$ because $\numBins<N$.
    When $\numBins \geq N$, apply \Cref{the:main}; it returns an optimal assignment, which is in particular an $(\mathrm{OPT}+1)$-approximation, in $\cO(N)$ time.
    The result follows.
\end{proof}

\subsection{Paper Organization and Overview of the Algorithm}

We prove \Cref{the:main} by designing and analyzing an algorithm to solve $\LBIF_{N\le \numBins}$ in $\cO(N)$ time and space. 
The overall algorithm works in several phases, each described as a simple algorithm. First, we check whether a solution with $\cOPT = 1$ exists. If so, we construct such an assignment. 
The characterization and algorithm for this case are deferred to \Cref{sec:OPT1}.
If no $\cOPT = 1$ solution exists, we apply the algorithm presented in \Cref{sec:OPT2}. This algorithm is guaranteed to produce a solution with $\cOPT \le 2$. Running these two phases in sequence guarantees an optimal solution.   
\Cref{sec:con} concludes the paper with some final remarks.

For the remainder of this paper, we assume $\lceil\totSize/\numBins \rceil > 1$. The special case $\lceil\totSize/\numBins \rceil = 1$ is trivial to solve by simply assigning one element of size $1$ to a total of $\totSize$ bins.

\section{Combinatorial Load at Most Two}\label{sec:OPT2}
In this section, we present an algorithm that \emph{always} \ie{for any input instance of $\LBIF_{N\le \numBins}$} produces an assignment with combinatorial load at most $2$.
We first present the algorithm in \Cref{sec:alg-at-most-2} and analyze the behavior in \Cref{sec:analysis-at-most-2}.

Let $\meanLoad \coloneqq \lceil \totSize/\numBins \rceil$ be the average \targetLoad\ per bin rounded up. By simple calculation, $t = \totSize - \numBins(\meanLoad-1) \in {\{1,\ldots,\numBins\}}$ is then the number of bins requiring \targetLoad\ $\meanLoad$.
Indeed, since $\meanLoad = \lceil \totSize/\numBins \rceil$, we have $\meanLoad-1 < \totSize/\numBins \le \meanLoad$,
which implies $0 < \totSize - \numBins(\meanLoad-1) \le \numBins$, so $t \in \{1,\ldots,\numBins\}$.
The multiset of strictly positive integers $\ites=\{\ite{1}, \ite{2}, \dots, \ite{N}\}$ initially consists of the sizes of the items to be assigned.
During the algorithm, we assign items or fragments of items from $\ites$ to bins and update each $\ite{i}$ to always represent the size of the fragment of item $i$ that is still to be assigned.
To facilitate the exposition, we define similarly $\capas=\{\capa{1}, \capa{2}, \dots, \capa{\numBins}\}$ as the multiset of $\numBins$ target bin loads. We remove elements from $\capas$ once their corresponding bins are completely filled.
We write $\capas_{\meanLoad-1}$ and $\capas_{\meanLoad}$ for the multisets consisting of elements in $\capas$ with value $(\meanLoad-1)$ and $\meanLoad$, respectively. 
Elements of $\capas_{\meanLoad-1}$ and elements of $\capas_{\meanLoad}$ are called $(\meanLoad-1)$-bins and $\meanLoad$-bins, respectively.
In the algorithm,
we only keep track of
their cardinalities $|\capas_{\meanLoad-1}|= \numBins-t$ and $|\capas_\meanLoad|= t = \totSize - \numBins(\meanLoad-1)$, respectively.
Therefore, updating these numbers is easily done with constant overhead. %

\subsection{The Algorithm}\label{sec:alg-at-most-2}
\paragraph{Overview.} The algorithm for  constructing a solution with combinatorial load at most $2$ is split into two parts. First, in \Cref{alg:s-less-c}, we iterate over the items that are guaranteed to fill any bin completely, possibly filling more than one bin at a time. Afterwards, \Cref{alg:s-is-c} fills the remaining bins with fragments from at most two items.
Full bins are removed from $\capas$, and $\ites$ is updated on the fly by removing the fragments of the items used.
\Cref{alg:s-less-c} is applied when $|\ites|<|\capas|$, whereas \Cref{alg:s-is-c} is applied when $|\ites|=|\capas|$.
We will show that we run \Cref{alg:s-less-c} until $|\ites|=|\capas|$, and from then on only run \Cref{alg:s-is-c}.

Inside the algorithms, we maintain a partition of $\ites$ into $\ites_{\mathrm{small}} = \{s \in \ites : s \le \meanLoad-1\}$ and
$\ites_{\mathrm{large}} = \{s \in \ites : s \ge \meanLoad\}$. The latter is the multiset of items that can each fill any bin completely.
Since all sizes are integers, every item belongs to exactly one of the two multisets; in particular, an item of size exactly $\meanLoad-1$ is small, even though it can fill an $(\meanLoad-1)$-bin on its own.
The multisets $\ites$, $\capas$, $\ites_\mathrm{small}$, and $\ites_\mathrm{large}$ will be updated in $\cO(1)$ time
as item fragments are consumed and produced.
We assume that values reaching zero are automatically removed from these multisets.

The formal problem definition represents an assignment by a matrix
$X=(x_{i,j})$. In the algorithms, we encode this matrix as a collection of
records, also denoted by $X$. Each record is a quadruple $(i,k,s,j)$, where
$k$ denotes the number of fragments of size $s$ from the item with index $i$,
and $j$ is either a bin index or the symbol $\bot$.
If $j$ is a bin index, then $k=1$, and the record assigns one fragment of
size $s$ from item $i$ to bin $j$. The fragment need not fill the bin by
itself. If several records have the same bin index $j$, their fragments are assigned to
the same bin, and their sizes together equal its \targetLoad.
If $j=\bot$, the record represents a batch of $k$ fragments of size $s$ from
item $i$.
Each fragment is placed in a distinct empty bin whose \targetLoad\ is $s$, such that
each of these bins is filled by a single fragment. The record assigns a total size of $k\cdot s$ from item $i$.
If $\ite{i}>k\cdot s$, the remaining size $\ite{i}-k\cdot s$ stays in $\ites$. The bin indices are omitted because bins
with the same \targetLoad\ are interchangeable.

Every record stored in $X$ has $k\geq1$. If the algorithm computes a
multiplicity of zero, no record is added. \Cref{alg:s-less-c} uses only
records with $j=\bot$, whereas \Cref{alg:s-is-c} uses explicit bin indices
to indicate which fragments are placed together.

\subsubsection{The Case $|\ites| < |\capas|$: Filling Empty Bins with Large Items}
\begin{algorithm}[t]
\DontPrintSemicolon
\KwIn{$\ites = \ites_{\mathrm{small}} \cup \ites_{\mathrm{large}}$, bin \targetLoads\ $\capas=\capas_{\meanLoad-1} \cup \capas_{\meanLoad}$, \targetLoad\ $\ell>1$}
$X \leftarrow \emptyset$\;
\While{$|\ites| < |\capas|$}{
Pick an item $\ite{i} \in \ites_{\mathrm{large}}$\;

\algstep{Step 1: fill entirely as many empty $(\meanLoad{-}1)$-bins as possible}
$a_i \leftarrow \min\!\bigl(\lfloor \ite{i} / (\meanLoad - 1) \rfloor,\; |\capas_{\meanLoad-1}|,\; |\capas|-|\ites| \bigr)$\;
$X \leftarrow X \cup \bigl\{(i,\, a_i,\, \meanLoad - 1,\, \bot) : a_i \neq 0\bigr\}$ \tcp*{$a_i$ $(\meanLoad{-}1)$-fragments of $\ite{i}$ into distinct $(\meanLoad{-}1)$-bins}
$\ite{i} \leftarrow \ite{i} - a_i (\meanLoad - 1)$;\quad remove $a_i$ bins from $\capas_{\meanLoad-1}$\;
\lIf{$\ite{i}=0$}{remove $\ite{i}$ from $\ites$ \textbf{else} reclassify $\ite{i}$  into $\ites_{\mathrm{small}}$ or $\ites_{\mathrm{large}}$}
\algstep{Step 2: fill entirely as many empty $\meanLoad$-bins as possible}
$b_i \leftarrow \min\!\bigl(\lfloor \ite{i} / \meanLoad \rfloor,\; |\capas_{\meanLoad}|,\; |\capas|-|\ites| \bigr)$\;
$X \leftarrow X \cup \bigl\{(i,\, b_i,\, \meanLoad,\, \bot) : b_i \neq 0 \bigr\}$ \tcp*{$b_i$ $\meanLoad$-fragments of $\ite{i}$ into distinct $\meanLoad$-bins}
$\ite{i} \leftarrow \ite{i} - b_i \cdot \meanLoad$;\quad remove $b_i$ bins from $\capas_{\meanLoad}$\;
\lIf{$\ite{i}=0$}{remove $\ite{i}$ from $\ites$ \textbf{else} reclassify $\ite{i}$  into $\ites_{\mathrm{small}}$ or $\ites_{\mathrm{large}}$}
}
\Return $X$\;
\caption{The case $|\ites| < |\capas|$: fill empty bins with whole fragments of a large item.\label{alg:s-less-c}}
\end{algorithm}

To keep the running time bounded by $\cO(N)$ rather than $\cO(B)$, each
iteration of the \textbf{while} loop in \Cref{alg:s-less-c} processes one
large item. In that iteration, the item may be divided into fragments that
fill several bins.
Rather than recording a separate assignment for every bin, the algorithm adds at most two
records to $X$: one representing all $(\meanLoad-1)$-bins filled from the selected item
and one representing all $\meanLoad$-bins filled from it.

In each iteration, \Cref{alg:s-less-c} selects an item from
$\ites_{\mathrm{large}}$. It first fills as many $(\meanLoad-1)$-bins as
possible. Once no such bins remain, it proceeds to fill the $\meanLoad$-bins.
Concretely, Step~1 assigns $a_i$ fragments of size $\meanLoad-1$ of item $i$ to as many $(\meanLoad-1)$-bins, and Step~2 then assigns $b_i$ fragments of size $\meanLoad$ of the residual item to as many $\meanLoad$-bins, where
\begin{align*}
    a_i &= \min\bigl(\lfloor \ite{i} / (\meanLoad - 1) \rfloor,\; |\capas_{\meanLoad-1}|,\; |\capas|-|\ites| \bigr), &
    b_i &= \min\bigl(\lfloor \ite{i} / \meanLoad \rfloor,\; |\capas_{\meanLoad}|,\; |\capas|-|\ites| \bigr),
\end{align*}
and $\ite{i}$, $\capas$, and $\ites$ in the definition of $b_i$ refer to the values after Step~1.

The first term in each of the definitions of $a_i$ and $b_i$ limits the
assignment according to the remaining size of item $i$. The second limits it
according to the number of available bins of the relevant \targetLoad. The
third, $|\capas|-|\ites|$, prevents the number of remaining bins from falling
below the number of remaining items. Each positive value $a_i$ or $b_i$ is
stored in a single record, irrespective of how many bins it represents.

As we will argue in \Cref{obs:alg1-terminates}, an item in
$\ites_{\mathrm{large}}$ is available as long as $|\ites|<|\capas|$. 
Hence, each iteration fills at least one bin and removes at most one item. The
algorithm continues until $|\ites|=|\capas|$, at which point
\Cref{alg:s-is-c} is applied to the remaining items and bins, together with
the partial assignment $X$.

\subsubsection{The Case $|\ites| = |\capas|$: Filling a Bin with One Small Item and (a Fragment of) One Large Item}
Once we have reached the point where $|\ites| = |\capas|$, we continue to fill the bins one by one. In broad terms, in each iteration of \Cref{alg:s-is-c}, exactly one bin and one item in $\ites_{\mathrm{small}}$
will be removed by filling the bin with the item in 
$\ites_{\mathrm{small}}$ complemented (if required) with a fragment of an item in $\ites_{\mathrm{large}}$.
More precisely, an iteration picks any $\iteSmall \in \itesSmall$, any $\iteLarge \in \itesLarge$, and any remaining bin with \targetLoad\ $\capa{j}$, places $\iteSmall$ into bin $j$, and fills the residual capacity $R \coloneqq \capa{j} - \iteSmall$ with a fragment of size $R$ of $\iteLarge$.
Once $\ites_{\mathrm{small}}$ or $\ites_{\mathrm{large}}$ are empty, we can continue with a simple matching of items to bins (\Cref{obs:empty-equality}). By this process, we can ensure that at most two items are assigned to any bin.  

\begin{algorithm}[t]
\DontPrintSemicolon
\KwIn{$\ites = \ites_{\mathrm{small}} \cup \ites_{\mathrm{large}}$, \targetLoads\ $\capas$ with $|\ites| = |\capas|$, target load $\meanLoad>1$, partial assignment $X$}
\While{$\itesSmall \neq \emptyset$ \textbf{and} $\itesLarge \neq \emptyset$}{

\algstep{Step 1: pick a small item, a large item, and a bin}
Pick any $\iteSmall \in \ites_{\mathrm{small}}$  with index $i_\smallrm$, any $\iteLarge \in \ites_{\mathrm{large}}$ with index $i_\largerm$, and any $\capa{j} \in \capas$\;
$R \leftarrow \capa{j} - \iteSmall$ \tcp*{$0 \le R \le \meanLoad - 1 < \iteLarge$}

\algstep{Step 2: pack bin $j$ with $\iteSmall$ and an $R$-fragment of $\iteLarge$}
$X \leftarrow X \cup \bigl\{(i_\smallrm,\,1,\,\iteSmall,\, j)\bigr\} \cup \bigl\{(i_\largerm,\,1,\,R,\, j) : R \neq 0 \bigr\}$ 
\;
Remove $\capa{j}$ from $\capas$;\quad remove $\iteSmall$ from $\ites$\;

\algstep{Step 3: update the residual large item}
Replace $\iteLarge$ by $(\iteLarge - R)$ in $\ites$\;
Reclassify the residual into $\ites_{\mathrm{small}}$ or $\ites_{\mathrm{large}}$\;
}
\While{$\ites \neq \emptyset$}{
\algstep{Step 4: once $\itesSmall$ or $\itesLarge$ is empty, match the elements exactly}

   Pick any $\ite{i}\in \ites$ with index $i$, and any $c_j\in \capas$\;
   $X \leftarrow X \cup \bigl\{(i,\,1,\,\ite{i},\, j)\bigr\}$ \tcp*{exact-fit deposits}
   Remove $c_j$ from $\capas$; remove $\ite{i}$ from $\ites$\;

}
\Return $X$\;
\caption{The case $|\ites| = |\capas|$: fill each bin with one small item and possibly part of one large item.\label{alg:s-is-c}}
\end{algorithm}

Notably, while $\iteSmall$ is fully assigned and removed and $\capa{j}$ is fully filled and removed in \Cref{alg:s-is-c}, the item $\iteLarge$ will never be fully removed in Step 3.  Thus, $|\ites| = |\capas|$ is maintained.
\subsection{Analysis}\label{sec:analysis-at-most-2}
We will now show that \Cref{alg:s-less-c} and \Cref{alg:s-is-c} together do in fact compute a solution to any instance of \LBIF with $|\ites| \le |\capas|$ with combinatorial load of at most $2$.
The heart of our correctness proof is to show that the following invariants are maintained at every step:

\begin{invariant}
Before and after every iteration of \Cref{alg:s-less-c,alg:s-is-c},
 \label{inv:size}
\begin{align}
    \sum_{\ite{i} \in \ites} \ite{i} &= \sum_{\capa{j} \in \capas} \capa{j}.
\end{align}
\end{invariant}

\begin{invariant}
 \label{inv:count}
 
Before and after every iteration of \Cref{alg:s-less-c,alg:s-is-c},
\begin{align}
    |\ites| &\le |\capas|.
\end{align}
\end{invariant}

\begin{lemma}
    \Cref{inv:size,inv:count} hold before the start of the algorithm.
\end{lemma}

\begin{proof}
By the definition of $\numBins=|\capas|$, $t$, and $\meanLoad$, we have $\sum_{\capa{j} \in \capas} \capa{j} = t\meanLoad + (\numBins-t)(\meanLoad-1) = \numBins(\meanLoad-1) + t = \totSize =  \sum_{\ite{i} \in \ites} \ite{i}$,
so~\Cref{inv:size} holds initially.
By the definition of the problem, \Cref{inv:count} also holds initially.
\end{proof}

\begin{lemma}\label{lem:algo1-inv}
    \Cref{alg:s-less-c} preserves \Cref{inv:size,inv:count}.
\end{lemma}

\begin{proof}
\textbf{\Cref{inv:size}:} Step 1 removes $a_i$ bins of \targetLoad\ $(\meanLoad-1)$ and reduces $\ite{i}$ by $a_i(\meanLoad-1)$. Step 2 removes $b_i$ bins of \targetLoad\ $\meanLoad$ and reduces $\ite{i}$ by $b_i \cdot \meanLoad$. In both steps, the total remaining item size and the total remaining \targetLoad decrease by the same amount, preserving~\Cref{inv:size}.
\textbf{\Cref{inv:count}:} The factor $|\capas| - |\ites|$ in the calculation of $a_i$ and $b_i$ in \Cref{alg:s-less-c} ensures that the case $|\ites| > |\capas|$ cannot occur. Thus, \Cref{inv:count} is preserved.

\end{proof}

With the following observation, we can see that \Cref{alg:s-less-c} must terminate with $|\ites| = |\capas|$. That is, while  $|\ites| < |\capas|$, there must be an element $\ite{}\in \itesLarge$ causing another iteration of \Cref{alg:s-less-c}.
\begin{observation}\label{obs:alg1-terminates}
    While $|\ites| < |\capas|$, $\itesLarge \ne \emptyset$.
\end{observation}

\begin{proof}
    Assume $\ites_{\mathrm{large}} = \emptyset$. Then every $s \in \ites$ satisfies $s \le \meanLoad-1$. Thus, since  $|\ites| < |\capas|$,
\[
    \sum_{s \in \ites} s \le |\ites|(\meanLoad-1)
    < |\capas|(\meanLoad-1) \le \sum_{\capa{j} \in \capas} \capa{j},
\]
which contradicts~\Cref{inv:size}. Hence, $\ites_{\mathrm{large}} \neq \emptyset$.
\end{proof}
The following observation justifies the use of Step 4 in \Cref{alg:s-is-c}.

\begin{observation}\label{obs:empty-equality}
While \Cref{inv:size}  holds, $|\ites| = |\capas|$, and either $\itesSmall = \emptyset$ or $\itesLarge = \emptyset$, then $\ites = \capas$ and $\ite{}=\capa{}$ for any pair of $\ite{}\in\ites$, $\capa{} \in\capas$.
\end{observation}
\begin{proof}
Assume $\itesSmall = \emptyset$. We will show by contradiction that $s = \meanLoad$ for any $s\in \itesLarge$. Assume there exists an $s\in \itesLarge$ with $s>\meanLoad$. We obtain
\[
\sum_{s \in \ites} s > |\ites|\cdot\meanLoad = |\capas|\cdot\meanLoad \ge \sum_{c \in \capas} c,
\]
which contradicts \Cref{inv:size}. Thus, $s = \meanLoad$ for all $s\in \itesLarge$. Now assume there exists a $c\in \capas$ with size $c \ne \meanLoad$. That is, $c = \meanLoad -1$. We obtain
\[
\sum_{s \in \ites} s = |\ites|\cdot\meanLoad = |\capas|\cdot\meanLoad > \sum_{c \in \capas} c,
\]
which again contradicts \Cref{inv:size}, proving our observation for the case $\itesSmall = \emptyset$. 
The argument for $\itesLarge = \emptyset$ is analogous.
\end{proof}

We use \Cref{obs:empty-equality} to prove the validity of the invariants for \Cref{alg:s-is-c} in the following lemma.
\begin{lemma}\label{lem:algo2-inv}
    \Cref{alg:s-is-c} preserves \Cref{inv:size,inv:count}. 
\end{lemma}

\begin{proof}
\textbf{\Cref{inv:size}:} Steps 2 and 3 of \Cref{alg:s-is-c} remove a bin of \targetLoad\ $\capa{j}$ from $\capas$, and item fragments of total size $\iteSmall + R = \iteSmall + \capa{j} - \iteSmall = \capa{j}$ from $\ites$. 
Thus, each iteration of the first while loop removes equal total values from $\capas$ and $\ites$, preserving \Cref{inv:size}. Consequently, the invariant holds upon termination of the first while loop. By \Cref{obs:empty-equality}, the first execution of Step 4 therefore removes identical values from $\capas$ and $\ites$, preserving \Cref{inv:size}. The same reasoning applies inductively to every subsequent execution of Step 4.

\textbf{\Cref{inv:count}:} Step 2 removes one item from $\ites$ and one bin from $\capas$. 
The item $i_\largerm$ will not be removed, as for the remaining size, $(\iteLarge - R)$, we have: $(\iteLarge - R) = \iteLarge - \capa{j} + \iteSmall \ge \meanLoad - \meanLoad + 1 = 1$.
Step 4 trivially removes one item and one bin per iteration.
Thus, \Cref{inv:count} is preserved in either case.

\end{proof}

We observe that once we switch from \Cref{alg:s-less-c} to \Cref{alg:s-is-c}, we will remain in \Cref{alg:s-is-c} until the end as in each iteration of \Cref{alg:s-is-c}, we either solve the problem completely or remove exactly one element from $\ites$ and one from $\capas$.\\

\begin{lemma}\label{lem:correct1}
  \Cref{alg:s-less-c} and \Cref{alg:s-is-c} terminate with a valid assignment.
\end{lemma}

\begin{proof}
  This follows directly from the invariants, 
  \Cref{obs:alg1-terminates,obs:empty-equality}
  and from the fact that, in every iteration, at least one bin is completely filled.
\end{proof}

\begin{lemma}\label{lem:correct2}
In \Cref{alg:s-less-c} and \Cref{alg:s-is-c}, any bin is filled with at most 2 item fragments in total.
\end{lemma}
\begin{proof}
Each bin is always assigned only one item (in \Cref{alg:s-less-c} or Step 4 of \Cref{alg:s-is-c}) or one small item and part of one large item (in Step 2 of \Cref{alg:s-is-c}).
\end{proof}

\begin{lemma}\label{lem:OPT2}
    For any $\LBIF_{N\le \numBins}$ and $\meanLoad > 1$, \Cref{alg:s-less-c} and \Cref{alg:s-is-c} construct an assignment with combinatorial load at most $2$ in $\cO(N)$ time using $\cO(N)$ space. 
\end{lemma}

\begin{proof}
The overarching algorithm consists of two phases: \Cref{alg:s-less-c} (while $|\ites| < |\capas|$) and \Cref{alg:s-is-c} (once $|\ites| = |\capas|$). All operations in both phases trivially take $\cO(1)$ time. 
In each iteration of \Cref{alg:s-less-c}, the cardinality of $\itesLarge$ is reduced by one or the algorithm terminates.
Thus \Cref{alg:s-less-c}, goes through at most $|\itesLarge|$ iterations. 
In \Cref{alg:s-is-c}, in each iteration of either loop, the cardinality  of $\ites$ is reduced by one.
Thus, the total number of iterations in \Cref{alg:s-less-c} and \Cref{alg:s-is-c} together is bounded by $|\itesLarge| + N \leq 2N$, yielding $\cO(N)$ total time.

The only non-constant space requirement comes from the assignment $X$. Since at most two quadruples are added to $X$ during any iteration of either algorithm, and as argued before, the overall algorithm has $\cO(N)$ iterations, the algorithm uses $\cO(N)$ space.

The validity of the resulting assignment from the entire algorithm and its combinatorial load follow from \Cref{lem:correct1,lem:correct2}. 
\end{proof}

\section{Combinatorial Load Exactly One}\label{sec:OPT1}

In the previous section, we presented an algorithm that finds an assignment with combinatorial load at most $2$, for any $\LBIF_{N\le \numBins}$ instance.
This is sufficient for \Cref{cor:strong-pol}, i.e., providing a strongly polynomial-time $(\mathrm{OPT}+1)$-approximation algorithm for \LBIF (regardless of $N$ and $\numBins$).
However, our ultimate goal is \Cref{the:main}: an exact algorithm for $\LBIF_{N\le \numBins}$. 
Since \Cref{sec:OPT2} focused on providing an assignment with combinatorial load at most $2$, all that remains is to provide a method for detecting whether an assignment with combinatorial load exactly $1$ exists, and if so, constructing one. 
Recall the notation $\meanLoad \coloneqq \lceil \totSize/\numBins \rceil$ for the \targetLoad\ rounded up and $t \coloneqq \totSize - \numBins(\meanLoad-1)$ for the number of bins with \targetLoad\ $\meanLoad$.
As before, the trivial case $\meanLoad\le 1$ is omitted.

The following observation is crucial for the design of our algorithm.

\begin{observation}
For any $\LBIF_{N\le \numBins}$ instance with 
$\cOPT=1$, 
the size $\ite{i}$ of each item must be representable as
\[
    \ite{i} = k_i \meanLoad + h_i(\meanLoad-1),
\]
where $k_i, h_i \in \mathbb{N}_{\ge 0}$ are the number of bins of \targetLoad\ $\meanLoad$ and $(\meanLoad-1)$ filled by fragments from $\ite{i}$, respectively. Furthermore, the number $h_i$ of $(\meanLoad-1)$-fragments must be of the form
\[
    h_i = \meanLoad \cdot m_i + f_i, \qquad \text{where } f_i \coloneqq (-\ite{i}) \bmod \meanLoad,
\]
for some integer $m_i \in \mathbb{N}_{\ge 0}$.
\label{obs:opt1}
\end{observation}
\begin{proof}
The first part, $\ite{i} = k_i \meanLoad + h_i(\meanLoad-1)$, is trivial, as for any assignment with combinatorial load $1$, any item must be assigned exactly to a combination of bins (each with \targetLoad\ $(\meanLoad-1)$ or $\meanLoad$).
For the second part of the observation, broadly speaking, the argument boils down to the fact that after assigning all $(\meanLoad-1)$-sized fragments of $\ite{i}$ to bins of \targetLoad\ $(\meanLoad-1)$, any remainder of $\ite{i}$, though possibly a remainder of $0$, 
must be divisible by $\meanLoad$ in order to fill up bins with \targetLoad $\meanLoad$ exactly.

Let us first consider the minimal number of $(\meanLoad-1)$-sized fragments an item $\ite{i}$ must contain in its decomposition  into fragments such that the item size is divisible by $\meanLoad$; i.e., the smallest number $f_i$ such that $(\ite{i}-f_i(\meanLoad-1))\bmod \meanLoad = 0$.
Choosing $f_i = (-\ite{i}) \bmod \meanLoad$, we obtain:

\begin{eqnarray*}
    \left(\ite{i}-f_i(\meanLoad-1)\right)\bmod \meanLoad    
    &=& \left( \ite{i}-(( - \ite{i} ) \bmod \meanLoad) (\meanLoad-1)  \right)\bmod \meanLoad  \\  
    &=& \left( \ite{i} - \meanLoad((-\ite{i})\bmod \meanLoad) + ((-\ite{i})\bmod \meanLoad ) \right) \bmod \meanLoad \\
    &=& \left( \ite{i} + ((-\ite{i}) \bmod \meanLoad) \right) \bmod \meanLoad \\
    &=& \left( \ite{i} - \ite{i} \right) \bmod \meanLoad
    = 0
\end{eqnarray*}

By the definition of $f_i$, we have $0\le f_i<\meanLoad$. Thus, $f_i = (- \ite{i} ) \bmod \meanLoad$ is the smallest non-negative number that satisfies $(\ite{i}-f_i(\meanLoad-1))\bmod \meanLoad = 0$.
For the term $\meanLoad\cdot m_i$, we analyze the remaining item size after $f_i$ fragments of size  $(\meanLoad-1)$ have been assigned to as many $(\meanLoad-1)$-bins and removed.
Let the remaining size to be distributed from item $\ite{i}$ be denoted by $r_i$. 
That is, 
$r_i = \ite{i} - f_i(\meanLoad-1) = k_i \meanLoad + h_i(\meanLoad-1) - f_i(\meanLoad-1)  = k_i \meanLoad + g_i(\meanLoad-1)$, where $g_i = h_i - f_i$.
By the argument above, $r_i \bmod \meanLoad = 0$. Trivially, $k_i \meanLoad \bmod \meanLoad = 0$. Hence, $g_i(\meanLoad-1) \bmod \meanLoad = 0$ must hold as well. Since $(\meanLoad-1)$ and $\meanLoad$ are coprime and thus, do not share a prime factor, it must be that $g_i \bmod \meanLoad = 0$ for any $\cOPT=1$ solution and can thus be written in the form $g_i = \meanLoad \cdot m_i$.
\end{proof}

Based on \Cref{obs:opt1}, we describe an algorithm (see \Cref{alg:load-1} for pseudocode) that detects whether $\cOPT=1$ and, if so, produces such an assignment. The algorithm proceeds in three steps, which we detail next.
It uses the following quantities.
For each item $i$, $f_i = (-\ite{i}) \bmod \meanLoad$ is the number of \emph{forced} fragments of size $\meanLoad-1$ from \Cref{obs:opt1}, and $F \coloneqq \sum_{i=1}^N f_i$ is their total number.
The counter $r$ is the number of $(\meanLoad-1)$-bins that are still empty after the forced fragments have been placed; initially, $r = \numBins - t - F$.
Finally, $n_i$ is the number of blocks of $\meanLoad$ fragments of size $\meanLoad-1$ each that Step~2 takes from item $i$; it plays the role of $m_i$ in \Cref{obs:opt1}.

\begin{algorithm}[t]
\DontPrintSemicolon
\KwIn{$\ites=\{\ite{1},\dots,\ite{N}\}$, bin \targetLoads\ $\capas$, target load $\meanLoad>1$}
$X \leftarrow \emptyset$\;
\algstep{Step 1: forced $(\meanLoad{-}1)$-fragments}
\ForEach{$\ite{i} \in \ites$}{
  $f_i \leftarrow (-\ite{i}) \bmod \meanLoad$\;
  \lIf{$\ite{i} < f_i\,(\meanLoad-1)$}{\Return \textsc{Fail}}
  $X \leftarrow X \cup \{(i, f_i, \meanLoad-1, \bot): f_i > 0\}$ \tcp*{$f_i$ $(\meanLoad-1)$-fragments of $\ite{i}$ into distinct $(\meanLoad-1)$-bins}  $\ite{i} \leftarrow \ite{i} - f_i\,(\meanLoad-1)$\;
}
$F \leftarrow \sum_i f_i$\;
\lIf{$F > \numBins - t$}{\Return \textsc{Fail}}
\lIf{$(\numBins - t - F) \bmod \meanLoad \neq 0$}{\Return \textsc{Fail}}
\BlankLine
\algstep{Step 2: bulk-fill remaining $(\meanLoad{-}1)$-bins in blocks of $\meanLoad$ fragments}
$r \leftarrow \numBins - t - F$\;
\ForEach{$\ite{i} \in \ites$ \textnormal{with} $\ite{i} \ge \meanLoad(\meanLoad-1)$ \KwSty{while} $r > 0$}{
  $n_i \leftarrow \min\!\big(\lfloor \ite{i} / (\meanLoad\,(\meanLoad-1)) \rfloor,\; r/\meanLoad\big)$\;
  $X \leftarrow X \cup \{(i, n_i \cdot \meanLoad, \meanLoad-1, \bot): n_i>0\}$ \tcp*{$n_i \cdot \meanLoad$ $(\meanLoad-1)$-fragments of $\ite{i}$ into distinct $(\meanLoad{-}1)$-bins}
  $\ite{i} \leftarrow \ite{i} - n_i\,\meanLoad\,(\meanLoad-1)$\;
  $r \leftarrow r - n_i\, \meanLoad$\;
}
\lIf{$r > 0$}{\Return \textsc{Fail}}
\BlankLine
\algstep{Step 3: fill the $\meanLoad$-bins}
\ForEach{$\ite{i} \in \ites$ \textnormal{with} $\ite{i} > 0$}{
  $X \leftarrow X \cup \{(i, \ite{i} / \meanLoad, \meanLoad, \bot)\}$ \tcp*{$(\ite{i} / \meanLoad)$ $\meanLoad$-fragments of $\ite{i}$ into distinct $\meanLoad$-bins}
}
\Return $X$\;
\caption{Combinatorial load of $1$: Test whether an assignment with combinatorial load $1$ exists, and if so, construct such assignment by decomposing the items into $\numBins - t$ and $t=\totSize - \numBins(\meanLoad-1)$ fragments of size $(\meanLoad-1)$ and $\meanLoad$, resp.\label{alg:load-1}}
\end{algorithm}

Step 1 iterates over all items $\ite{i}$ in $\ites$, computing
$f_i$. 
If $\ite{i}<f_i(\meanLoad-1)$, the algorithm correctly returns \textsc{Fail}. 
If $\ite{i} \geq f_i (\meanLoad -1)$, we assign these $f_i$ fragments to $f_i$ empty bins of \targetLoad\ $(\meanLoad-1)$, each fragment filling one bin entirely. We deduct $f_i(\meanLoad-1)$ from $\ite{i}$. By construction, the updated size $\ite{i}$ is now non-negative and divisible by $\meanLoad$. 

The algorithm computes the total number $F=\sum_{\ite{i}\in \ites}f_i$. Recall that $\numBins-t$ is the number of available bins with \targetLoad\ $(\meanLoad-1)$. If $F>\numBins-t$, a valid $\cOPT=1$ packing is impossible, and the algorithm correctly returns \textsc{Fail}. Otherwise, after depositing the $F$ forced fragments, $\numBins - t - F$ empty bins of \targetLoad\ $(\meanLoad-1)$ remain. 

By \Cref{obs:opt1}, any additional fragments of size $(\meanLoad-1)$ extracted from an item must be extracted in blocks of $\meanLoad$ fragments 
in order to preserve the item's divisibility by $\meanLoad$. 
Let us call such blocks \emph{full blocks}. Therefore, the total number of remaining $(\meanLoad-1)$-bins across the entire instance must be a multiple of $\meanLoad$. If it is not, the algorithm correctly returns \textsc{Fail}. If it is, then we proceed to Step 2.

Step 2 greedily fills the remaining $r=\numBins-t-F$ bins with target load $(\meanLoad-1)$ using full blocks. As verified, $r$ is a multiple of $\meanLoad$. The algorithm iterates through the remaining available items. The loop terminates early when all $(\meanLoad-1)$-bins are filled.

We compute the number $n_i$ of full
blocks that can be extracted from the remaining item size $\ite{i}$, i.e., after having extracted the $f_i$ fragments of size $(\meanLoad-1)$. Since each full block has total size $\meanLoad(\meanLoad-1)$, no more than $\lfloor (\ite{i} / (\meanLoad\,(\meanLoad-1)) \rfloor$
blocks can be extracted from $\ite{i}$. Moreover, in Step 2 we only fill 
the $r$ as of yet remaining empty $(\meanLoad-1)$-bins, 
and each block fills $\meanLoad$ of them. Thus, no more than $r/\meanLoad$ blocks are needed. Hence, $n_i\leq r/\meanLoad$.

In each iteration with $\ite{i}$, we take the minimum of these two values as $n_i$ and extract exactly $n_i \cdot \meanLoad$ fragments of size $(\meanLoad-1)$ from  item $\ite{i}$. 
These are assigned to the available
distinct $(\meanLoad-1)$-bins.
Then we deduct the deposited size from the item
and decrease the number of remaining empty $(\meanLoad-1)$-bins by the exact number of bins we just filled.
After iterating through all items, if $r > 0$, it means we exhausted all available item sizes
but failed to fill all the $(\meanLoad-1)$-bins, in which case the algorithm returns \textsc{Fail}. Otherwise, it proceeds to Step 3.

In Step 3, all $(\meanLoad-1)$-bins have already been filled.
Furthermore, the remaining size of every item $\ite{i}$ is guaranteed to be a multiple of $\meanLoad$. Therefore, it can be sliced into an exact integer number of $\meanLoad$-fragments, each of which we assign to a yet-empty $\meanLoad$-bin, thereby filling the $t$ bins with target load $\meanLoad$. 

\begin{lemma}\label{lem:OPT1}
    For any $\LBIF_{N\le \numBins}$ instance with $\meanLoad>1$, \Cref{alg:load-1} checks
    whether $\cOPT=1$ in $\cO(N)$ time.
    If $\cOPT=1$, an assignment with this value is constructed in $\cO(N)$ time using $\cO(N)$ space.
\end{lemma}

\begin{proof}
We first prove that \Cref{alg:load-1} returns a valid assignment with combinatorial load $1$ if and only if such an assignment exists.

$(\Rightarrow)$ We show that the algorithm returns a valid $\cOPT=1$ assignment.

By construction, the algorithm assigns fragments of size $(\meanLoad-1)$ only to $(\meanLoad-1)$-bins (Steps 1 and 2), and fragments of size $\meanLoad$ only to $\meanLoad$-bins (Step 3). In Step 1, $f_i$ fragments are extracted from $\ite{i}$. The condition $\ite{i} \ge f_i(\meanLoad-1)$ guarantees that $\ite{i}$ is large enough to extract all $f_i$ fragments from it. In Step 2, $n_i$ full blocks of size $\meanLoad(\meanLoad-1)$ are extracted. The condition $n_i \le \lfloor \ite{i} / (\meanLoad(\meanLoad-1)) \rfloor$ similarly guarantees that, after removing the $f_i$ forced fragments, $\ite{i}$ is large enough to contain $n_i$ blocks. Thus, all fragment sizes are valid. Furthermore, because $r=0$ at the end of Step 2, all $(\numBins-t)$ bins of \targetLoad\ $(\meanLoad-1)$ are completely filled. At the start of Step 3, the remaining size $\ite{i}$ of every item is divisible by $\meanLoad$ because, after subtracting the $f_i$ forced fragments to make the remainder divisible by $\meanLoad$ (Step 1), we have only subtracted multiples of $\meanLoad$ (Step 2). Therefore, Step 3 partitions the remaining item sizes into fragments of size $\meanLoad$. Since the total size $\totSize = t \meanLoad + (\numBins-t)(\meanLoad-1)$ is preserved and the $(\meanLoad-1)$-bins are full, the remaining total item size fills the remaining $t$ bins of \targetLoad\ $\meanLoad$. Thus, the maximum combinatorial load is exactly 1.

$(\Leftarrow)$ If an $\cOPT=1$ assignment exists, we must show the algorithm does not return \textsc{Fail}. 

By \Cref{obs:opt1}, any valid $\cOPT=1$ assignment must allocate exactly $h_i = \meanLoad \cdot m_i + f_i$ fragments of size $(\meanLoad-1)$ from item $\ite{i}$. Therefore, every item $\ite{i}$ must have a total size of at least $f_i(\meanLoad-1)$; otherwise, no such assignment could exist. Consequently, the check $\ite{i} < f_i(\meanLoad-1)$ in Step 1 will correctly pass. The total number of forced fragments of size $(\meanLoad-1)$, $F = \sum f_i$, represents the minimum number of $(\meanLoad-1)$-bins required. If $F > (\numBins-t)$, no valid assignment can exist, so the algorithm passes this check. Furthermore, by \Cref{obs:opt1}, the remaining $(\numBins-t-F)$ bins of \targetLoad\ $(\meanLoad-1)$ must be filled in blocks of $\meanLoad$ of them to preserve the divisibility by $\meanLoad$ of the items. This implies that $(\numBins-t-F)$ must be divisible by $\meanLoad$. Thus, the check $(\numBins-t-F) \bmod \meanLoad \neq 0$ will also pass. 

In Step 2, because $r$ is a multiple of $\meanLoad$ and $\cOPT=1$ guarantees  
that the number of fragments of size $(\meanLoad-1)$ is sufficient to fill all remaining empty bins with target load $(\meanLoad-1)$, the full-block filling strategy fills all these bins. Thus, the algorithm will not fail in Step 2. 

Finally, we analyze the complexity. Each item is processed a constant number of times, and all arithmetic operations are constant-time. Clearly, the calculation of line $F \leftarrow \sum_i f_i$ is dominated by the preceding loop. Therefore, the total running time is $\cO(N)$. The output assignment stores only one quadruple per deposited block, so it uses $\cO(N)$ space.
\end{proof}

The correctness of \Cref{alg:load-1} directly yields the following characterization.

\begin{corollary}\label{lem:OPT=1}
For any $\LBIF_{N\leq \numBins}$ instance,
\begin{eqnarray}\cOPT=1& \Leftrightarrow
&\forall i \in \{1, \dots, N\}, \ \ite{i} \geq  ((-\ite{i})\bmod \meanLoad)(\meanLoad-1) \\ &{\rm and}& \ 
\sum_{i=1}^N (-\ite{i})\bmod \meanLoad \leq \numBins\meanLoad - \totSize\\
&{\rm and}& \
\sum_{i=1}^N \lfloor \frac{\ite{i}- ((-\ite{i})\bmod \meanLoad)(\meanLoad-1)}{\meanLoad(\meanLoad-1)}
\rfloor \cdot \meanLoad
\geq \numBins\meanLoad - \totSize
- \sum_{i=1}^N ( - \ite{i})\bmod \meanLoad \\ &{\rm and}& \
\left( \numBins\meanLoad - \totSize
- \sum_{i=1}^N (-\ite{i})\bmod \meanLoad \right)\bmod \meanLoad=0.
\end{eqnarray}
Here, $\numBins\meanLoad - \totSize = \numBins - t$ is the number of $(\meanLoad-1)$-bins.
\end{corollary}

\Cref{lem:OPT2,lem:OPT1} prove \Cref{the:main}: we first run \Cref{alg:load-1} on a copy of the item sizes. If it returns an assignment, this assignment has combinatorial load $1$ and is optimal. Otherwise, $\cOPT \ge 2$ and we run \Cref{alg:s-less-c} followed by \Cref{alg:s-is-c} on the original item sizes, which by \Cref{lem:OPT2} yields combinatorial load at most $2$, which is again optimal. Both phases take $\cO(N)$ time and space.

\section{Concluding Remarks}\label{sec:con}

The main contributions of this paper are twofold. 

From the results perspective, \Cref{the:main} and \Cref{cor:strong-pol} improve on the widely used algorithm of Kobert et al.~\cite{DBLP:conf/wabi/KobertFAS14}, and they may inspire improvements in other settings in parallel and distributed systems or in iterative algorithms that rebalance their workload at every step. In particular, \Cref{the:main} establishes that the otherwise NP-hard problem has a linear-time solution under assumptions that naturally arise in high-performance computing. Furthermore, the simplicity of the presented algorithm makes it readily applicable within existing scheduling frameworks. 

From an educational perspective, our paper may serve as material for a class on mathematical optimization, showcasing a remarkably simple and optimal algorithm obtained through combinatorial insight. In particular, our algorithm makes no use of advanced data structures: it only uses standard arithmetic and set operations, making it especially suitable as teaching material since students can focus on the underlying ideas rather than on implementation details.

\section*{Declarations}
We used an LLM to produce the Ti$k$Z code for \Cref{fig:example-load2} based on a textual description we provided.
Furthermore, LLMs generated feedback on the stylistic polishing of the final manuscript, which we then implemented ourselves where we thought it was useful.
All technical content and ideas were developed and written by us, the human authors.
We assume responsibility for all content.

\bibliographystyle{alphaurl}
\bibliography{sample}

\end{document}